%% file: main.tex
\documentclass[a4paper]{article}

\usepackage[english]{babel}
\usepackage[utf8x]{inputenc}
\usepackage[T1]{fontenc}

\usepackage[a4paper,top=2.54cm,bottom=2.54cm,left=2.54cm,right=2.54cm,marginparwidth=1.75cm]{geometry}

\usepackage{amsmath}
\usepackage{graphicx}
\usepackage[colorinlistoftodos]{todonotes}
\usepackage[colorlinks=false, allcolors=blue]{hyperref}
\usepackage{parskip}
\usepackage{setspace}

\usepackage{cite}

\usepackage{amsmath}

\usepackage{graphicx}

\usepackage{multirow}
\usepackage{float}
\usepackage{etoolbox}
\usepackage[font=large,labelfont=bf]{caption}
\usepackage{mwe}

\title{\textbf{Pressure-dependent mechanical and thermodynamic properties of newly discovered cubic Na\textsubscript{2}He}}

\author{\normalsize Md. Zahidur Rahaman\textsuperscript{1}\\\vspace{.01in}
\\\small \textit{Department of Physics}\\\small \textit{Pabna University of Science and Technology, Pabna-6600, Bangladesh}\\\small \textit{Bangladesh University of Engineering and Technology, Dhaka, Bangladesh}\\\small \textit{\href{mailto:zahidur.physics@gmail.com}{zahidur.physics@gmail.com}}\\\vspace{.01in}\\
\normalsize Md. Lokman Ali\textsuperscript{2}\\\vspace{.01in}
\\\small \textit{Department of Physics}\\\small \textit{Pabna University of Science and Technology, Pabna-6600, Bangladesh}\\\small \textit{\href{mailto:lokman.cu12@gmail.com}{lokman.cu12@gmail.com}}\\\vspace{.01in}\\
\normalsize Md. Atikur Rahman\textsuperscript{3*}\\\vspace{.01in}
\\\small \textit{Department of Physics}\\\small \textit{Pabna University of Science and Technology, Pabna-6600, Bangladesh}\\\small \textit{\href{mailto:atik0707phy@gmail.com}{atik0707phy@gmail.com}}
\thanks{Corresponding author}
}

\date{\small (09 July, 2017)}

\begin{document}
\maketitle

\input{abstract}
\input{introduction}
\input{method}  
\input{result}

\input{conclusion}

\input{acknowledge}

\input{reference}

\input{figures}
\input{table}

\end{document}

%% file: abstract.tex
\begin{abstract}
\addcontentsline{toc}{section}{Abstract}
\normalsize
\singlespace{
Recently for the first time, a stable compound of He and Na
(Na\textsubscript{2}He) is predicted at high pressure. We explore the
pressure-dependent elastic, mechanical and thermodynamic properties of
this newly discovered Na\textsubscript{2}He by using \emph{ab initio}
technique. The calculation presents good accordance between the
theoretical and experimental lattice parameters. Though the most stable
structure of Na\textsubscript{2}He is found at 300 GPa, present study
ensures the mechanical stability of this compound up to 500 GPa. The
study of Cauchy pressure, Pugh's ratio, and Poisson's ratio implies the
ductile manner of Na\textsubscript{2}He up to 500 GPa. According to the
value of Poisson's ratio the bonding force exists in
Na\textsubscript{2}He is central. The study of Zener anisotropy factor
indicates that Na\textsubscript{2}He is an anisotropic material but near
at 300 GPa approximately isotropic nature of Na\textsubscript{2}He is
revealed. The study of the bulk modulus, shear modulus, Young's modulus
and Vickers hardness implies that the hardness of Na\textsubscript{2}He
can be improved by applying external pressure. However, the Debye
temperature, melting temperature and minimum thermal conductivity of
Na\textsubscript{2}He are also calculated and discussed at different
pressures.

\textbf{\small{Keywords:}} Na\textsubscript{2}He, Structural properties,
Mechanical properties, Thermodynamic properties.}

\end{abstract}

%% file: introduction.tex
\clearpage
\section*{I. Introduction}
\addcontentsline{toc}{section}{Introduction}
\large 
\doublespacing

Helium (He) is chemically inert because of its closed-shell electronic
configuration. For this reason, helium is very unwilling to form any
stable compound in normal condition. However, helium possesses many
attractive chemical properties such as it has highest ionization
potential (24.59 eV) \cite{1} and zero electron affinity \cite{2}. On the
other hand, helium is available in large quantities in the universe. For
this reason, many scientists have tried to discover a stable compound of
helium in recent decades. For the first time in 2015, Xiao Dong \emph{et
al} predicted a stable compound of sodium (Na) and helium (He) at
pressure 300 GPa \cite{3}. Though the predicted compound
Na\textsubscript{2}He is stable above 113 GPa. They carried out both
experimental and theoretical investigation for ensuring the dynamical
stability of Na\textsubscript{2}He. However, the mechanical stability of
Na\textsubscript{2}He is still vague. They also showed that
Na\textsubscript{2}He is not an inclusion material. The formation of
Na\textsubscript{2}He is strongly exothermic and a wide bandgap appears when helium is inserted into sodium \cite{3}. They also studied
the electronic structure and pressure-dependent bandgap of
Na\textsubscript{2}He.

It is well known that pressure has a significant effect on the physical
properties as well as chemical properties of materials. When a material
is compressed it is generally deformed. This deformation behavior helps
to understand the nature of solid-state theories \cite{4}. High pressure
can also lead to the phase transition of a material \cite{5, 6, 7}. Therefore, it is very important to investigate the pressure effect on the mechanical and thermodynamic properties of newly reported Na\textsubscript{2}He. In this paper, we study the elastic, mechanical and thermodynamic properties of Na\textsubscript{2}He by using the DFT (Density Functional Theory) based first principles code CASTEP under different pressure up to 500 GPa. In this present letter, we also ensure the mechanical stability of Na\textsubscript{2}He by theoretical means.

%% file: method.tex
\section*{II. Method of computation}
\addcontentsline{toc}{section}{Method of computation}
\large 
\doublespacing

The first principles calculations were performed in the CASTEP computer
code with generalized gradient approximation (GGA) using the PBE
(Perdew, Burke, and Ernzerhof) exchange-correlation functional
\cite{8, 9, 10, 11, 12}. The structural relaxation was obtained by using the BFGS minimization scheme \cite{13}. In this case, the convergence criteria were set to 0\emph{.}5 × 10\textsuperscript{−5} eV\emph{/}atom for energy,
0.05 GPa for stress and 0.03 eV/ Å for the force. The energy cutoff was set
at 1000 eV with 10×12×10 grids in primitive cells of
Na\textsubscript{2}He through the investigation. Stress-strain method
\cite{13} was used for obtaining the elastic stiffness constants at
different pressures. In this case, the maximum value of strain amplitude
was set to 0.003 through the whole elastic constant calculations at
various pressures.

%% file: result.tex
\section*{III. Results and discussion}
\addcontentsline{toc}{section}{Results and discussion}

\subsection*{A. Electronic structure}
\addcontentsline{toc}{subsection}{Electronic structure}
\large 
\doublespacing

The first stable compound of helium
Na\textsubscript{2}He belongs to \emph{Fm-3m} (225) space group with the
cubic crystal structure, which retains four formula unit with 12 atoms
per unit cell \cite{3}. The experimentally evaluated lattice constant of
Na\textsubscript{2}He is 3.95 Å at 300 GPa \cite{3}. By minimizing the
total energy the atomic positions and lattice parameters have been
optimized as a function of normal pressure. Since Na\textsubscript{2}He
is more stable at 300 GPa, the crystal structure treated with 300 GPa is
shown in Fig. \ref{fig:Fig. 1}. In the structure of Na\textsubscript{2}He 4\emph{a} atomic site is fully occupied by He atoms with fractional coordinates
(0, 0, 0) and 8\emph{c} atomic site is fully occupied by Na atoms with
fractional coordinates (0.25, 0.25, 0.25). The calculated structural
parameters of Na\textsubscript{2}He at 300 GPa are tabulated in Table \ref{table 1} along with experimental values. The evaluated lattice constant at 300
GPa in this present study shows minor deviation (0.12\%) from the
experimentally evaluated value bearing the reliability of the present
investigation.

To investigate the effect of pressure on the structure of
Na\textsubscript{2}He, the variations of unit cell volume and lattice
parameter of Na\textsubscript{2}He have been studied under different
pressure from 100 GPa to 500 GPa. Fig. \ref{fig:Fig. 2}(a) represents the change of
lattice constant and unit cell volume of Na\textsubscript{2}He with
external pressure. It is evident that the normalized lattice constant
and cell volume decreases with pressure. The compression of the crystal
is not uniform over the whole range of pressure. With the increase of
pressure, the relative compression of the structure is decreased due to
the repulsive interaction among atoms. The more the pressure is applied
the nearer the atoms are and hence their repulsive interaction is
strengthened, which leads to the above effect. Fig. \ref{fig:Fig. 2}(b) represents the
pressure-volume plot of Na\textsubscript{2}He from which we determine
the pressure derivative bulk modulus \emph{\(B^\prime\)} of Na\textsubscript{2}He listed in Table \ref{table 1}.


\subsection*{B. Mechanical properties}
\addcontentsline{toc}{subsection}{Mechanical properties}
\large 
\doublespacing

In order to justify the mechanical stability and determine the different mechanical properties of solids, the single crystal elastic constants have been calculated by linear fitting of the calculated stress-strain function \cite{14}. The cubic Na\textsubscript{2}He belongs to three independent elastic constants \emph{C\textsubscript{11}}, \emph{C\textsubscript{12}} and \emph{C\textsubscript{44}} . The evaluated elastic constants of Na\textsubscript{2}He at different pressure are tabulated in Table \ref{table 2}. As shown in Table \ref{table 2} it is clear that the studied compound is mechanically stable up to 500 GPa since the obtained values of \emph{C\textsubscript{ij}} obey the well established Born stability
criteria of solids \cite{15} given below.

\emph{C\textsubscript{11}} \textgreater{} 0, \emph{C\textsubscript{44}}
\textgreater{} 0, \emph{C\textsubscript{11}} --
\emph{C\textsubscript{12}} \textgreater{} 0 and
\emph{C\textsubscript{11}} + 2\emph{C\textsubscript{12}} \textgreater{}
0\\
The pressure-dependent elastic constants and different mechanical
properties of Na\textsubscript{2}He are demonstrated in Fig. \ref{fig:Fig. 3}. The
value of \emph{C\textsubscript{11}} implies the stiffness of compounds
against (100)\(\langle100\rangle\) uniaxial strain. We can see from Table \ref{table 2} and Fig. \ref{fig:Fig. 3}(a) that Na\textsubscript{2}He becomes stiffer with the increase of pressure. On the other hand, the value of \emph{C\textsubscript{12}} and
\emph{C\textsubscript{44}} demonstrates the elasticity in shape. One can
see from Fig. \ref{fig:Fig. 3}(a) that with the increase of pressure both the shear constants are increased implying that the compound becomes more resistant to shear deformation with pressure.

The Cauchy pressure is defined as \emph{C\textsubscript{12} --
C\textsubscript{44}} which can be used to explain the atomic bonding
character in a compound \cite{16}. A compound will be nonmetallic
(metallic) if \emph{C\textsubscript{12} -- C\textsubscript{44}} is
negative (positive) \cite{17}. From Table \ref{table 2} we see that the value of
\emph{C\textsubscript{12} -- C\textsubscript{44}} for
Na\textsubscript{2}He is positive over the pressure range studied
indicating the metallic nature of Na\textsubscript{2}He. However, the Cauchy pressure is increased with the increase in pressure {[}Fig. \ref{fig:Fig. 3}(b){]}. Furthermore, the negative (positive) value of \emph{C\textsubscript{12}
-- C\textsubscript{44}} implies the brittle (ductile) manner of a
compound \cite{18}. As shown in Table \ref{table 2}, Na\textsubscript{2}He is ductile
and hence able to be drawn out into a thin wire. From Fig. \ref{fig:Fig. 3}(b) it is
noticeable that the more the pressure is applied the more ductile the
compound is implying the increasing metallicity of Na\textsubscript{2}He
with pressure.

From the single crystal elastic constants, \emph{C\textsubscript{ij}} the polycrystalline elastic moduli such as shear modulus \emph{G} and bulk modulus \emph{B} can be obtained according to Voigt-Reuss-Hill (VRH) approximation \cite{19}. From the computed bulk modulus \emph{B} and shear modulus \emph{G} the Poisson's ratio \emph{\(\nu\)} and Young's modulus \emph{E} can be obtained by using the following relations \cite{20},
\begin{equation}
\tag{1}
\begin{split}
\nu = \ \frac{3B - 2G}{2(3B + G)}
\end{split}
\end{equation}
\begin{equation}
\tag{2}
\begin{split}
E = \ \frac{9\text{GB}}{3B + G}
\end{split}
\end{equation}
The computed values of \emph{B}, \emph{G}, \emph{\(\nu\)}, and \emph{E} under different pressure for Na\textsubscript{2}He are listed in Table \ref{table 3}. It can be noted from Table \ref{table 3} that the computed value of the bulk modulus
\emph{B} matches well with that of obtained from fitting the third order
Birch-Murnaghan Equation of State at 300 GPa. The bulk modulus \emph{B}
is a measure of the ability of solids to defend compression. From Table \ref{table 3} we see that at 300 GPa (more stable structure) the value of \emph{B} for Na\textsubscript{2}He is 758.43 GPa which is comparatively so high than that of the well known hardest material diamond (\emph{B} = 443
GPa) \cite{21}. This high value of bulk modulus is indicating strong
chemical bonding in Na\textsubscript{2}He. A hard material possesses
high bonding covalency. In Na\textsubscript{2}He direct covalent
interaction is found for Na-Na \cite{3}, showing good consistency with
this study. With the increase in pressure, the bulk modulus also
increases {[}Fig. \ref{fig:Fig. 3}(c){]} indicating that the ability of the resist
deformation of Na\textsubscript{2}He is enhanced with pressure. The
shear modulus \emph{G} is a measure of the ability of solids to defend
plastic deformation. The value of \emph{G} depends upon the value of
\emph{C\textsubscript{44}}. The large value of
\emph{C\textsubscript{44}} and \emph{G} at 300 GPa (see Table \ref{table 2} and Table \ref{table 3}) implies a high level of resistant of Na\textsubscript{2}He against the shear strain. It can also be noticed from Fig. \ref{fig:Fig. 3}(c) that with
external pressure the value of \emph{G} is increased indicating the
enhancement of the ability to resist shear deformation with the pressure of
Na\textsubscript{2}He. The Young's modulus \emph{E} is a measure of the
ability of solids to defend longitudinal stress. It is also used to
judge the stiffness of a material. From Table \ref{table 3} it can be seen that at 300 GPa the value of \emph{E} is quite high for Na\textsubscript{2}He and with the increase in pressure \emph{E} increases gradually indicating the enhancement of stiffness of Na\textsubscript{2}He with external pressure.

Pugh's ratio usually the ratio between bulk modulus and shear modulus
(\emph{B}/\emph{G}) is a very useful index to detect the brittleness and
ductility of materials \cite{22}. \emph{B/G} \textless{} 1.75 implies the
brittle manner of a material and the value greater than 1.75 implies the
ductile manner of a compound. From Table \ref{table 3} one can see that
Na\textsubscript{2}He is ductile in nature shows good accordance with
the result having from the analysis of Cauchy pressure. The pressure-dependent Pugh's ratio of Na\textsubscript{2}He is shown in Fig. \ref{fig:Fig. 3}(d). It can be noted that at 100 GPa the fabrication of this compound is
easier. The Poisson's ratio \emph{\(\nu\)} is another useful index to explain the plasticity and bonding nature of solids. For the covalent crystal, the value of \emph{\(\nu\)} is 0.10 and 0.25 for ionic crystal. The value from 0.25 to 0.50 implies the central force in a solid \cite{23}. However, \emph{\(\nu\)} = 0.33 indicates metallic material \cite{24}. From Table \ref{table 3} we notice that all over the pressure range the Poisson's ratio is nearly similar and the value is around 0.39 implying the existence of the central force in Na\textsubscript{2}He. The pressure-dependent Poisson's ratio of Na\textsubscript{2}He is shown in Fig. \ref{fig:Fig. 3}(e). We see a similar trend between the pressure-dependent curve of \emph{B}/\emph{G} and \emph{\(\nu\)}.
Poisson's ratio is also used to predict the brittleness and ductility of
solids. If \emph{\(\nu\)} \textless{} 0.26 for a compound then the compound will be brittle in the manner otherwise the compound will be ductile in the manner \cite{25, 26}. According to this condition, Na\textsubscript{2}He is ductile in nature, accord well with other predictions.

To measure the rate of anisotropy in crystal the Zener anisotropy factor
\emph{A} is a very useful index which is given as follows \cite{28},
\begin{equation}
\tag{3}
\begin{split}
A = \ \frac{2C_{44}}{(C_{11} - C_{12})}
\end{split}
\end{equation}
For the complete isotropic crystal, the value of \emph{A} is unity. The value greater or smaller than unity implies the rate of anisotropy in solid. The calculated values of \emph{A} at various pressures are listed in
Table \ref{table 3}. As shown in Table \ref{table 3} the anisotropic nature of Na\textsubscript{2}He has revealed. The pressure-dependent behavior of
\emph{A} is illustrated in Fig. \ref{fig:Fig. 3}(f). With pressure up to 300 GPa the rate of anisotropy is decreased but after that, a strange behavior is noticed. More theoretical work is required to understand this mysterious behavior of \emph{A} for Na\textsubscript{2}He.

The Vickers hardness \emph{H\textsubscript{v}} is a very important
property of compound used to measure the hardness of solids. Chen
\emph{et al.} developed a very useful formula to predict the hardness of
solids given below \cite{27},
\begin{equation}
\tag{4}
\begin{split}
H_{V} = 2\left( K^{2}G \right)^{0.585} - \ 3
\end{split}
\end{equation}
Where, \emph{K} = \emph{G}/\emph{B}. The values of
\emph{H\textsubscript{v}} computed from Eq. 4 under different pressures
of Na\textsubscript{2}He are tabulated in Table \ref{table 3}. Evidently
\emph{H\textsubscript{v}} of Na\textsubscript{2}He increases gradually
with applied pressure {[}Fig. \ref{fig:Fig. 3}(g){]} implying that pressure has a significant effect on the hardness of Na\textsubscript{2}He.


\subsection*{C. Thermodynamic properties}
\addcontentsline{toc}{subsection}{Thermodynamic properties}
\large 
\doublespacing

Debye temperature \emph{\(\Theta\)\textsubscript{D}} is a very crucial
thermodynamic parameter which is related to directly or indirectly many
important thermal properties of solids such as specific heat of solid,
melting point, thermal expansion etc. It is the temperature which is
related to the highest normal mode of vibration of a crystal \cite{29}.
The Debye temperature can be obtained from different estimations. In
this study, we have used the calculated elastic constants to compute the
value of \emph{\(\Theta\)\textsubscript{D}} for Na\textsubscript{2}He. In this procedure, \emph{\(\Theta\)\textsubscript{D}} can be computed directly by using the following equation \cite{30},
\begin{equation}
\tag{5}
\begin{split}
\Theta_{D} = \ \frac{h}{k_{B}}\left( \frac{3N}{4\pi V} \right)^{\frac{1}{3}}{\times~ v}_{m}
\end{split}
\end{equation}
Where \emph{v\textsubscript{m}} is the average sound velocity which can
be calculated,
\begin{equation}
\tag{6}
\begin{split}
v_{m} = \ \left\lbrack \frac{1}{3}\left( \frac{2}{{v_{t}}^{3}} + \frac{1}{{v_{l}}^{3}} \right) \right\rbrack^{- \frac{1}{3}}
\end{split}
\end{equation}
Where \emph{v\textsubscript{l}} and \emph{v\textsubscript{t}} stands
for longitudinal wave velocity and transverse wave velocity respectively
which can be obtained as follows,
\begin{equation}
\tag{7}
\begin{split}
v_{l} = \left( \frac{3B + 4G}{3\rho} \right)^{\frac{1}{2}}
\end{split}
\end{equation}
And
\begin{equation}
\tag{8}
\begin{split}
v_{t} = \left( \frac{G}{\rho} \right)^{\frac{1}{2}}
\end{split}
\end{equation}
The computed values of \emph{v\textsubscript{l}},
\emph{v\textsubscript{t}} , \emph{v\textsubscript{m}}, and
\emph{\(\Theta\)\textsubscript{D}} of Na\textsubscript{2}He at different
pressures are listed in Table \ref{table 4}. The pressure-dependent behavior of \emph{\(\Theta\)\textsubscript{D}} for Na\textsubscript{2}He is illustrated in Fig. \ref{fig:Fig. 4}(a). It can be noticed from Fig. \ref{fig:Fig. 4}(a) that with the increase in pressure \emph{\(\Theta\)\textsubscript{D}} of Na\textsubscript{2}He is increased gradually.

The thermal conductivity of the material is defined as the capacity to
conduct heat energy through solid. The minimum thermal conductivity is a
crucial thermodynamic parameter. With the increase in temperature, the
thermal conductivity of solid is decreased and reaches a certain value
generally defined as the minimum thermal conductivity \cite{31}. The
minimum thermal conductivity \emph{K\textsubscript{min}} can be obtained
by using the following relation \cite{32},
\begin{equation}
\tag{9}
\begin{split}
K_{\min} = \ K_{B}v_{m}\ \left( \frac{M}{\text{n\(\rho\)}N_{A}} \right)^{- \ 2/3}
\end{split}
\end{equation}
Where \emph{v\textsubscript{m}} stands for average sound velocity,
\emph{K\textsubscript{B}} is the Boltzmann constant, \emph{n} is the
number of atoms per molecule, \emph{M} is the molecular mass, and
\emph{N\textsubscript{A}} is the Avogadro's number. The calculated
values of \emph{K\textsubscript{min}} at different pressures by using
Eq. 9 are listed in Table \ref{table 5} and plotted in Fig. \ref{fig:Fig. 4}(b). It can be noted
from Fig. \ref{fig:Fig. 4}(b) that with the increase in pressure
\emph{K\textsubscript{min}} of Na\textsubscript{2}He is increased
gradually.

The melting temperature of solid is another important thermodynamic
parameter. The melting temperature \emph{T\textsubscript{m}} of cubic
crystal can be computed as follows \cite{33},
\begin{equation}
\tag{10}
\begin{split}
T_{m} = 553 + 5.91C_{11}\ \pm 300\ K
\end{split}
\end{equation}
The calculated values of the melting temperature for
Na\textsubscript{2}He at different pressures are listed in Table \ref{table 5} and
illustrated in Fig. \ref{fig:Fig. 4}(c). It can be seen from Fig. \ref{fig:Fig. 4}(c) that with the
increase in pressure \emph{T\textsubscript{m}} of Na\textsubscript{2}He
is increased gradually. At high temperature, the anharmonic effect of the
specific heat at constant volume is suppressed and close to a value
known as Dulong-Petit limit \cite{34}. The Dulong-Petit limit of solid
can be evaluated as follows \cite{34},
\begin{equation}
\tag{11}
\begin{split}
Dulong - Petit\ limit = 3nN_{A}K_{B}
\end{split}
\end{equation}
Where \emph{N\textsubscript{A}} is defined as the Avogadro's constant
and \emph{K\textsubscript{B}} is the Boltzmann constant. The calculated
Dulong-Petit limit of Na\textsubscript{2}He using Eq. 11 is listed in
Table \ref{table 5}.

%% file: conclusion.tex
\section*{IV. Conclusions}
\addcontentsline{toc}{section}{Conclusions}
\large 
\doublespacing

In summary, a comprehensive investigation on the mechanical and
thermodynamic properties of the first stable compound of helium and sodium
Na\textsubscript{2}He has been carried out systematically by the theoretical
method. The mechanical stability of cubic Na\textsubscript{2}He has been
ensured through theoretical verification up to 500 GPa. Various
mechanical properties (bulk modulus \emph{B}, shear modulus \emph{G},
Young's modulus \emph{E}, \emph{B/G} values, Poisson's ratio \emph{\(\nu\)},
anisotropy factor \emph{A} and Vickers hardness
\emph{H\textsubscript{v}} ) of Na\textsubscript{2}He under different
pressure have also been reported. The study of the bulk modulus, shear
modulus, Young's modulus and Vickers hardness implies that the hardness
of Na\textsubscript{2}He can be improved by applying external pressure.
The study of Zener anisotropy factor indicates that
Na\textsubscript{2}He is an anisotropic material but near at 300 GPa
approximately isotropic nature of Na\textsubscript{2}He is revealed. The
study of Cauchy pressure, Pugh's ratio, and Poisson's ratio implies the
ductile manner of Na\textsubscript{2}He up to 500 GPa. The pressure-dependent behaviors of various thermal properties of
Na\textsubscript{2}He have also been reported and discussed. We hope the
predicted properties will have a good impact on comprehending the
mechanical response of newly discovered Na\textsubscript{2}He.

%% file: acknowledge.tex
\section*{Acknowledgments} 
\addcontentsline{toc}{section}{Acknowledgments}
\large 
\doublespacing

We thank Professor Dr. A.K.M. Akther Hossain for valuable discussions and crucial suggestions. We would also like to thank Department of Physics, Pabna University of Science and Technology, Bangladesh, for the laboratory support.

%% file: reference.tex
\begin{center}
\begin{tikzpicture}
\draw (-2,0) -- (2,0);
\filldraw [black,very thick] (0,0) (-1,0) -- (1,0);
\filldraw [black,ultra thick] (0,0) (-.5,0) -- (.5,0);
\end{tikzpicture}
\end{center}

\renewcommand{\section}[2]{}%

%% file: figures.tex
\clearpage
\begin{figure}[H]
\centering
\includegraphics[width=1\textwidth]{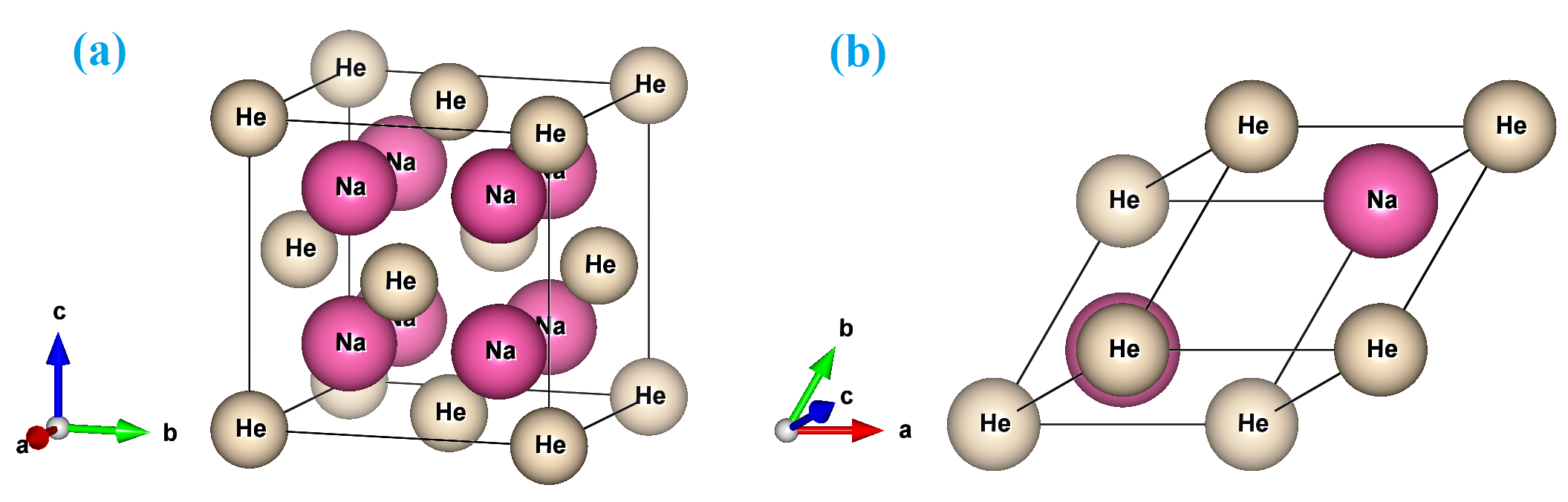}
\caption{Crystal structure of Na\textsubscript{2}He: (a) conventional unit cell and (b) primitive cell.} 
\label{fig:Fig. 1}
\end{figure}

\begin{figure}[H]
\centering
\includegraphics[width=1\textwidth]{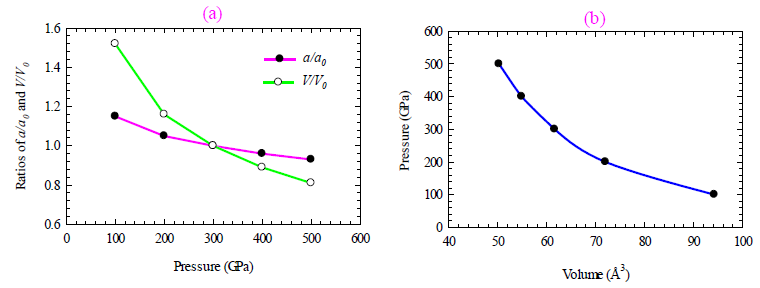}
\caption{(a) Change of lattice parameter and cell volume as a function of pressure (b) Birch-Murnaghan equation of state for Na\textsubscript{2}He.}  
\label{fig:Fig. 2}
\end{figure}

\begin{figure}[H]
\centering
\includegraphics[width=1\textwidth]{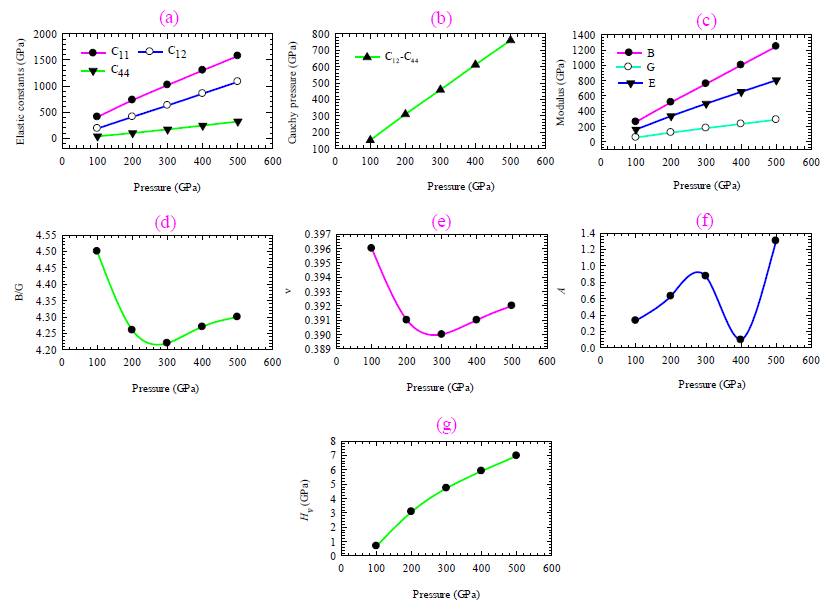}
\caption{Pressure-dependent mechanical properties of Na\textsubscript{2}He.} \label{fig:Fig. 3}
\end{figure}

\begin{figure}[H]
\centering
\includegraphics[width=1\textwidth]{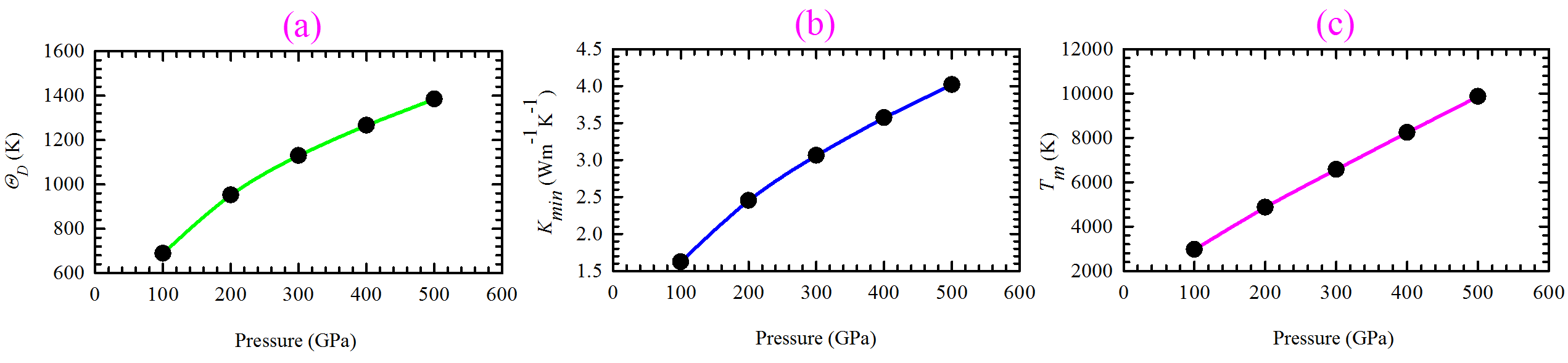}
\caption{Pressure-dependent thermodynamic properties of Na\textsubscript{2}He.}
\label{fig:Fig. 4}
\end{figure}

%% file: table.tex
\clearpage

\begin{table*}[ht]
\doublespacing
\large
\centering 

\caption{\large Calculated equilibrium Lattice constant ``\emph{a}'',
unit cell volume ``\emph{V}'', bulk modulus ``\emph{B}'' and its first
pressure derivative ``\emph{\(B^\prime\)}'' of Na\textsubscript{2}He at 300 GPa.}
\label{table 1}
\bigskip

\begin{tabular}[]{@{}cccc@{}}
\hline
Properties & Expt.\cite{3} & This study & Deviation from Expt.
(\%)\tabularnewline
\hline
\emph{a} (Å ) & 3.950 & 3.955 & 0.12\tabularnewline
\emph{V} (Å\textsuperscript{3}) & - & 61.864 & -\tabularnewline
\emph{B} (GPa) & - & 747.32 & -\tabularnewline
\emph{\(B^\prime\)} & - & 2.51 & -\tabularnewline
\hline
\end{tabular}

\end{table*}

\begin{table*}[ht]
\doublespacing
\large
\centering 

\caption{\large Evaluated elastic constants \emph{C\textsubscript{ij}}
(GPa) and Cauchy pressure (\emph{C\textsubscript{12}} --
\emph{C\textsubscript{44}} ) of Na\textsubscript{2}He under pressure.}
\label{table 2}
\bigskip

\begin{tabular}[]{@{}ccccc@{}}
\hline
\emph{P (GPa)} & \emph{C\textsubscript{11}} & \emph{C\textsubscript{12}}
& \emph{C\textsubscript{44}} & \emph{C\textsubscript{12} --
C\textsubscript{44}}\tabularnewline
\hline
100 & 407.05 & 187.55 & 36.54 & 151.01\tabularnewline
200 & 729.83 & 409.82 & 100.58 & 309.24\tabularnewline
300 & 1018.20 & 628.55 & 170.20 & 458.35\tabularnewline
400 & 1299.87 & 854.93 & 243.62 & 611.31\tabularnewline
500 & 1575.16 & 1081.55 & 321.77 & 759.78\tabularnewline
\hline
\end{tabular}

\end{table*}

\begin{table*}[ht]
\doublespacing
\large
\centering 

\caption{\large The evaluated bulk modulus \emph{B} (GPa), shear
modulus \emph{G} (GPa), Young's modulus \emph{E} (GPa), \emph{B/G}
values, Poisson's ratio \emph{\(\nu\)}, anisotropy factor \emph{A} and Vickers hardness \emph{H\textsubscript{v}} (GPa) of Na\textsubscript{2}He under different pressures.}
\label{table 3}
\bigskip

\begin{tabular}[]{@{}cccccccc@{}}
\hline
\emph{P (GPa)} & \emph{B} & \emph{G} & \emph{E} & \emph{B/G} & \emph{\(\nu\)}
& \emph{A} & \emph{H\textsubscript{v}}\tabularnewline
\hline
100 & 260.71 & 57.82 & 161.51 & 4.50 & 0.396 & 0.332 &
0.68\tabularnewline
200 & 516.49 & 121.23 & 337.29 & 4.26 & 0.391 & 0.628 &
3.07\tabularnewline
300 & 758.43 & 179.65 & 499.51 & 4.22 & 0.390 & 0.873 &
4.72\tabularnewline
400 & 1003.24 & 234.92 & 653.73 & 4.27 & 0.391 & 0.095 &
5.91\tabularnewline
500 & 1246.08 & 289.34 & 805.66 & 4.30 & 0.392 & 1.303 &
6.97\tabularnewline
\hline
\end{tabular}

\end{table*}

\begin{table*}[ht]
\doublespacing
\large
\centering 

\caption{\large The calculated density \emph{\(\rho\)} (in
gm/cm\textsuperscript{3}), transverse (\emph{v\textsubscript{t}}),
longitudinal (\emph{v\textsubscript{l}}), and average sound velocity
\emph{v\textsubscript{m}} (m/s) and Debye temperature
\emph{\emph{\(\Theta\)}\textsubscript{D}} (K) of Na\textsubscript{2}He.}
\label{table 4}
\bigskip

\begin{tabular}[]{@{}cccccc@{}}
\hline
Pressure (GPa) & \emph{\(\rho\)} & \emph{v\textsubscript{t}} & \emph{v\textsubscript{l}} & \emph{v\textsubscript{m}} & \emph{\emph{\(\Theta\)}\textsubscript{D}} \tabularnewline
\hline
100 & 3.51 & 4058.68 & 9810.21 & 4592.46 & 688.69\tabularnewline
200 & 4.60 & 5133.64 & 12141.64 & 5804.34 & 952.05\tabularnewline
300 & 5.37 & 5783.97 & 13632.33 & 6538.80 & 1129.59\tabularnewline
400 & 6.04 & 6236.50 & 14763.40 & 7051.52 & 1266.20\tabularnewline
500 & 6.60 & 6621.13 & 15724.26 & 7487.27 & 1384.54\tabularnewline
\hline
\end{tabular}

\end{table*}

\begin{table*}[ht]
\doublespacing
\large
\centering 

\caption{\large Calculated minimum thermal conductivity,
\emph{K\textsubscript{min}} (in Wm\textsuperscript{-1}K\textsuperscript{-1}), melting temperature,
\emph{T\textsubscript{m}} (in K) and the Dulong-Petit limit (in
J/mole.K) of Na\textsubscript{2}He at different pressures.}
\label{table 5}
\bigskip

\begin{tabular}[]{@{}cccc@{}}
\hline
Pressure (GPa) & \emph{K\textsubscript{min}} & \emph{T\textsubscript{m}} & Dulong-Petit limit \tabularnewline
\hline
100 & 1.62 & 2958.66 ± 300 &\tabularnewline
200 & 2.45 & 4866.29 ± 300 &\tabularnewline
300 & 3.06 & 6570.56 ± 300 & 74.80\tabularnewline
400 & 3.57 & 8235.23 ± 300 &\tabularnewline
500 & 4.02 & 9862.19 ± 300 &\tabularnewline
\hline
\end{tabular}

\end{table*}